# An Improvement of RC4 Cipher Using Vigenère Cipher

SEIFEDINE KADRY, MOHAMAD SMAILI

Lebanese University - Faculty of Science, Lebanon

*Address*: LIU, P.O. Box 5, Jeb Janeen, Khyara, Bekaa, Lebanon

E-mail: skadry@gmail.com

## Abstract

This paper develops a new algorithm to improve the security of RC4. Given that RC4 cipher is widely used in the wireless communication and has some weaknesses in the security of RC4 cipher, our idea is based on the combination of the RC4 and the poly alphabetic cipher Vigenère to give a new and more secure algorithm which we called VRC4. In this technique the plain text is encrypted using the classic RC4 cipher then re-encrypt the resulted cipher text using Vigenère cipher to be a more secure cipher text. For simplicity, we have implemented our new algorithm in Java Script taking into consideration two factors: improvement of the security and the time complexity. To show the performance of the new algorithm, we have used the well known network cracking software KisMac.

## 1. Introduction

The increasing in the electronic communication demands more and more security on the exchange of the critical information. Cryptography now a day's get more and more importance to address this issue [1]. In the cryptography, the original text usually called "plain text" and the encoded or altered text is called "cipher text". The conversion from plain text to cipher text is called "encoding", "encrypting", or "enciphering", and the opposite operation is called "decoding", "decrypting", or "deciphering". The cryptography allows two people, Alice and Bob, to exchange a message in such a way that other people, Eve, cannot understand the message (fig. 1).

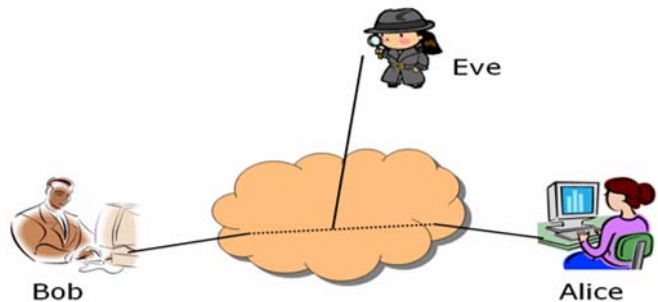

**Figure 1: Cryptography**

Encryption is mainly used to ensure secrecy. Companies usually encrypt their data before transmission to ensure that the data is secure during transmission. The encrypted data is sent over the public network and is decrypted by the intended recipient. Encryption works by running the data (represented as numbers) through a special encryption formula (called a key). Both the sender and the receiver know this key which may be used to encrypt and decrypt the data as shown in (fig. 2).

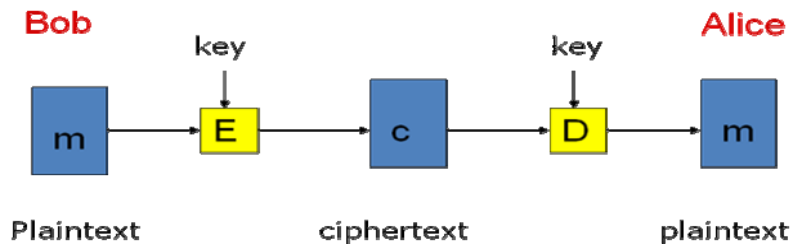

**Figure 2: Encryption/Decryption**

This paper is organized as follows: Section 2 presents the encryption process and the two categories of cipher Block and Stream. The RC4 cipher is explained in section 3. Section 4 talks about Vigenère, and section 5 explains in details the new proposed algorithm and the improvement of RC4 using Vigenère technology. The simulation and the evaluation of the new algorithm are given in Section 6. Finally, in section 7 we conclude the paper with future work.

## 2. Encryption Process

There are a variety of different types of encryption methods, they can be classified according to the way in which the plaintext is processed (which can be either stream cipher or block cipher), or according to the type of operations used for transforming plaintext to cipher text. The second class can be one of two styles, substitution [2] (which maps each element in the plaintext into another element) and transposition (which rearranges elements in the plaintext) .
Basically the two methods of producing cipher text are stream cipher and block cipher[3]. The two methods are similar except for the amount of data each encrypts on each pass. Most modern encryption schemes use some form of a block cipher.

## 2.1 Block and Stream Ciphers

Block and Stream Ciphers are two categories of ciphers used in classical cryptography [4, 5]. Block and Stream Ciphers differ in how large a piece of the message is processed in each encryption operation. Block ciphers encrypt plain text in chunks. Common block sizes are 64 and 128 bits. Stream ciphers encrypt plaintext one byte or one bit at a time. A stream cipher can be thought of as a block cipher with a really small block size. Generally speaking, block ciphers are more efficient for computers and stream ciphers are easier for humans to do by hand.

## 3. RC4

In cryptography, RC4 [6, 7] (also known as ARC4 or ARCFOUR meaning Alleged RC4) is the most widely-used software stream cipher and is used in popular protocols such as Secure Sockets Layer (SSL) (to protect Internet traffic) and WEP (to secure wireless networks). While remarkable for its simplicity and speed in software, RC4 has weaknesses that argue against its use in new systems. It is especially vulnerable when the beginning of the output key stream is not discarded, nonrandom or related keys are used, or a single key stream is used twice; some ways of using RC4 can lead to very insecure cryptosystems such as WEP.

RC4 was designed by Ron Rivest of RSA Security in 1987, and kept as a trade secret. It is officially termed "Rivest Cipher 4". The RC acronym is alternatively understood to stand for "Ron's Code. In September 1994, the RC4 algorithm was anonymously posted on the Internet on the Cipher punk's anonymous remailers list.

Several versions of the protocols are in wide-spread use in applications like web browsing, electronic mail, Internet faxing, instant messaging and voice-over-IP (VoIP).

The main factors which helped its deployment over such a wide range of applications consisted in its impressive speed and simplicity. Implementations in both software and hardware are very easy to develop.

## 3.1 RC4 Algorithm

The RC4 algorithm is remarkably simply and quite easy to explain. A variable-length key of from 1 to 256 bytes (8 to 2048 bits) is used to initialize a 256-byte state vector S, with elements S [0], S [1],..., S [255]. At all times, S contains a permutation of all 8-bit numbers from 0through 255. For encryption and decryption, a byte *k* is generated from S by selecting one of the 255 entries in a systematic fashion. As each value of *k* is generated, the entries in S are once again permuted [7]:

*Initialization of S*
To begin, the entries of S are set equal to the values from 0 through 255 in ascending order; that is; S[0] = 0, S[1] = 1,..., S[255] = 255. A temporary vector, T, is also created. If the length of the key K is 256 bytes, then K is transferred to T. Otherwise, for a key of length *keylen* bytes, the first *keylen* elements of T are copied from K and then K is repeated as many times as necessary to fill out T. These preliminary operations can be summarized as follows (fig. 3):

/* Initialization, */
for i = 0 to 255 do
S[i] = i;
T[i] = K [i mod keylen];
Next we use T to produce the initial permutation of S. This involves starting with S [0] and going through to S [255], and, for each S[i], swapping S[i] with another byte in S according to a scheme dictated by T[i]

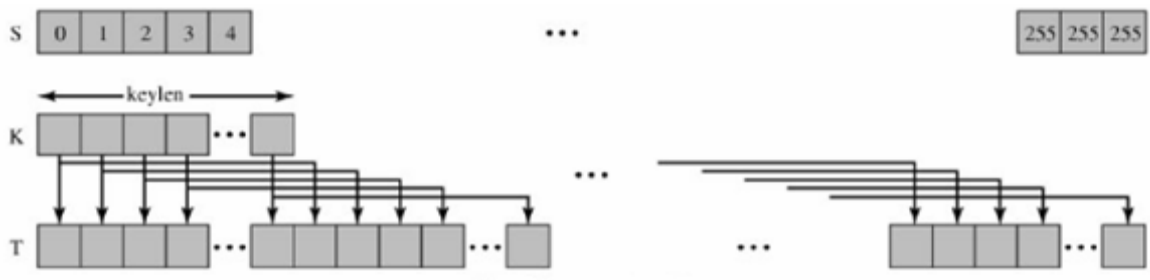

**Figure 3: Initialization step**

```
/* Initial Permutation of S */
j = 0;
for i = 0 to 255 do
j = (j + S[i] + T[i]) mod 256;
Swap (S[i], S[j]);
```

Because the only operation on S is a swap, the only effect is a permutation. S still contains all the numbers from 0 through 255 (fig. 4).

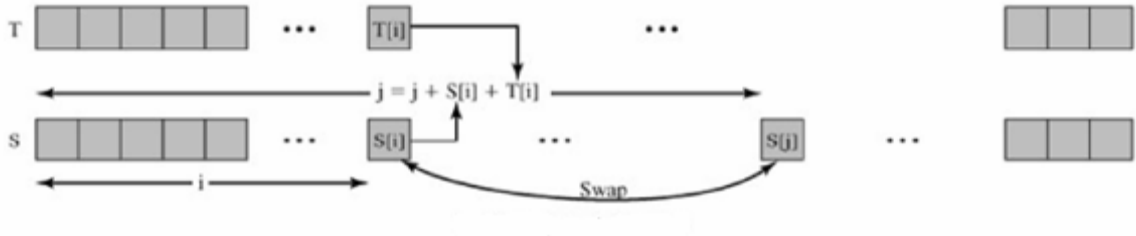

**Figure 4: Initial permutation of S**

*Stream Generation*

Once the S vector is initialized, the input key is no longer used. Stream generation involves cycling through all the elements of S[i], and, for each S[i], swapping S[i] with another byte in S according to a scheme dictated by the current configuration of S. After S [255] is reached, the process continues, starting over again at S [0], (fig. 5):

```
/* Stream Generation */
i, j = 0;
while (true)
i = (i + 1) mod 256;
j = (j + S[i]) mod 256;
Swap (S[i], S[j]);
t = (S[i] + S[j]) mod 256;
k = S[t];
```

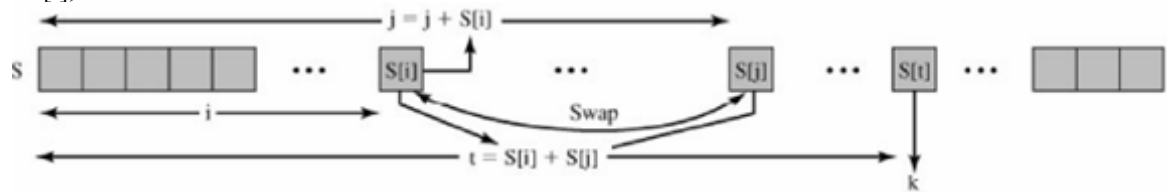

**Figure 5: Stream generation**

To encrypt, XOR the value *k* with the next byte of plaintext. To decrypt, XOR the value *k* with the next byte of cipher text (fig. 6).

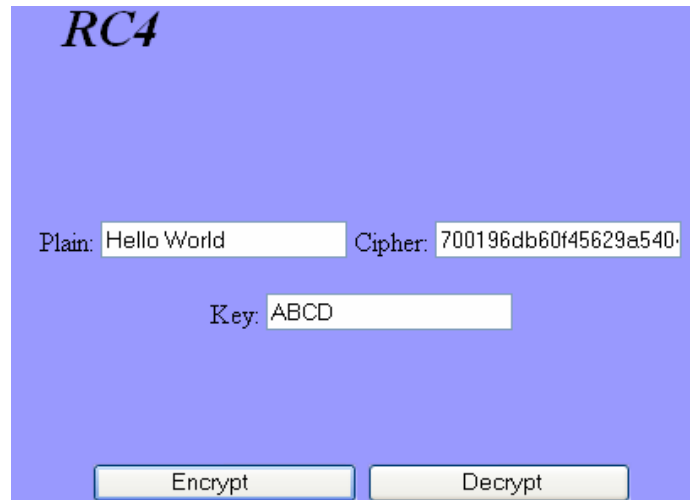
**Figure 6: Encryption/Decryption**

### 3.2 Strength of RC4

A number of papers have been published analyzing methods of attacking RC4 like [8-12]. None of these approaches is in practical against RC4 with a reasonable key length, such as 128 bits. The authors demonstrate that the WEP protocol, intended to provide confidentiality on 802.11 wireless LAN networks, is vulnerable to a particular attack approach. In essence, the problem is not with RC4 itself, but the way in which keys are generated for use as input to RC4. This particular problem does not appear to be relevant to other applications using RC4, and can be remedied in WEP by changing the way in which keys are generated. This problem points out the difficulty in designing a secure system that involves both cryptographic functions and protocols that make use of them.

For all stream ciphers, key lengths of up to and including 128 bits MUST be supported by the implementation, although any particular key may be shorter. Longer keys are strongly recommended.

RC4 allows for key sizes of up to 256 bytes. The key is used to permute the values of 0 to 255 in an array of 256 entries, thus creating the internal state for the pseudo random generator, whose output is used as one pad for encryption/decryption. The initial state after keying can thus be one out of 256! states, being equivalent to about 1676 bits Key state.

The algorithm is based on the use of a random permutation. The Problem is not with RC4 itself, but the way in which keys are generated for use as input to RC4.

### 4. Vigenère Cipher

The Vigenère cipher is a method of encrypting alphabetic text by using a series of different Caesar ciphers based on the letters of a keyword. It is a simple form of polyalphabetic substitution [13, 14].

The Vigenère Cipher has been reinvented many times. The method was originally described by Giovan Battista Bellaso in his 1553 book La cifra del. Sig. Giovan Battista Bellaso. However, the scheme was later misattributed to Blaise de Vigenère in the 19$^{th}$ century, and is now widely known as the "Vigenère cipher".

The Vigenère Cipher gained a reputation for being exceptionally strong. Noted author and mathematician Charles Lutwidge Dodgson (Lewis Carroll) called the Vigenère Cipher unbreakable in his 1868 piece "The Alphabet Cipher" in a children's magazine. In 1917, Scientific American described the Vigenère Cipher as "impossible of translation". This reputation was not deserved, since Kasiski entirely broke the cipher in the 19$^{th}$ century and some skilled cryptanalysts could occasionally break the cipher in the 16$^{th}$ century.

### 4.1 Vigenère Algorithm

The Vigenère square or Vigenère table, also known as the tabula recta, can be used for encryption and decryption. In a Caesar cipher, each letter of the alphabet is shifted along some number of places; for example, in a Caesar cipher of shift 3, A would become D, B would become E and so on. The Vigenère cipher consists of several Caesar ciphers in sequence with different shift values.

To encipher, a table of alphabets can be used, termed a tabula recta, Vigenère square, or Vigenère table. It consists of the alphabet written out 26 times in different rows, each alphabet shifted cyclically to the left compared to the previous alphabet, corresponding to the 26 possible Caesar ciphers. At different points in the encryption process, the cipher uses a different alphabet from one of the rows. The alphabet used at each point depends on a repeating keyword.

Suppose that the plaintext to be encrypted is: ATTACKATDAWN, the person sending the message chooses a keyword and repeats it until it matches the length of the plaintext, for example, the keyword "LEMON": LEMONLEMONLE, the first letter of the plaintext, A, is enciphered using the alphabet in row L, which is the first letter of the key. This is done by looking at the letter in row L and column A of the Vigenère square, namely L. Similarly, for the second letter of the plaintext, the second letter of the key is used. The letter at row E and column T is X. The rest of the plaintext is enciphered in a similar fashion [14]

Plaintext: ATTACKATDAWN
Key: LEMONLEMONLE
Cipher text: LXFOPVEFRNHR

Decryption is performed by finding the position of the cipher text letter in a row of the table, then taking the label of the column in which it appears as the plaintext. For example, in row L, the cipher text L appears in column A, which taken as the first plaintext letter. The second letter is decrypted by looking up X in row E of the table; it appears in column T, which is taken as the plaintext letter (fig. 7).

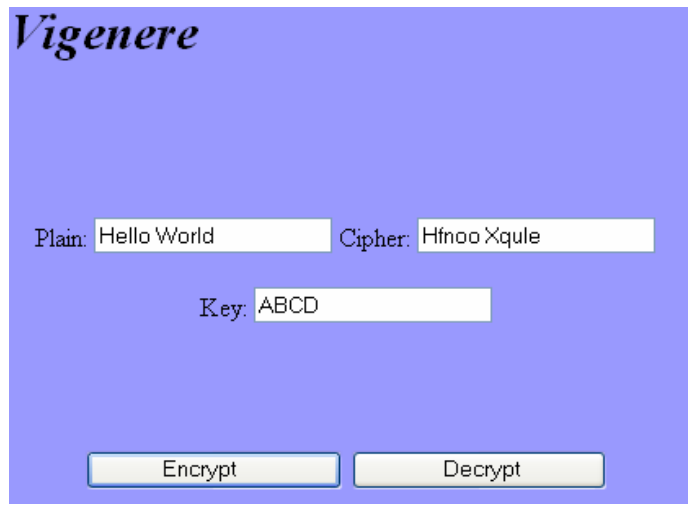

**Figure 7: Encryption/Decryption**

## 5. The Proposed Algorithm: VRC4

In our proposed algorithm, we have combined RC4 and Vigenère in the following fashion (fig. 10):

Step 1: Generate k using RC4
Step 2: J = random(0,255)
Step 3: C1 = Vigenère(Crc4[0:J], K[0:J])
       C2 = Vigenère(Crc4[J+1:length(Crc4)], K[J+1:255])
Step 4: C = C1 + C2 + J

Following the encryption/decryption process (fig. 8 and 9):

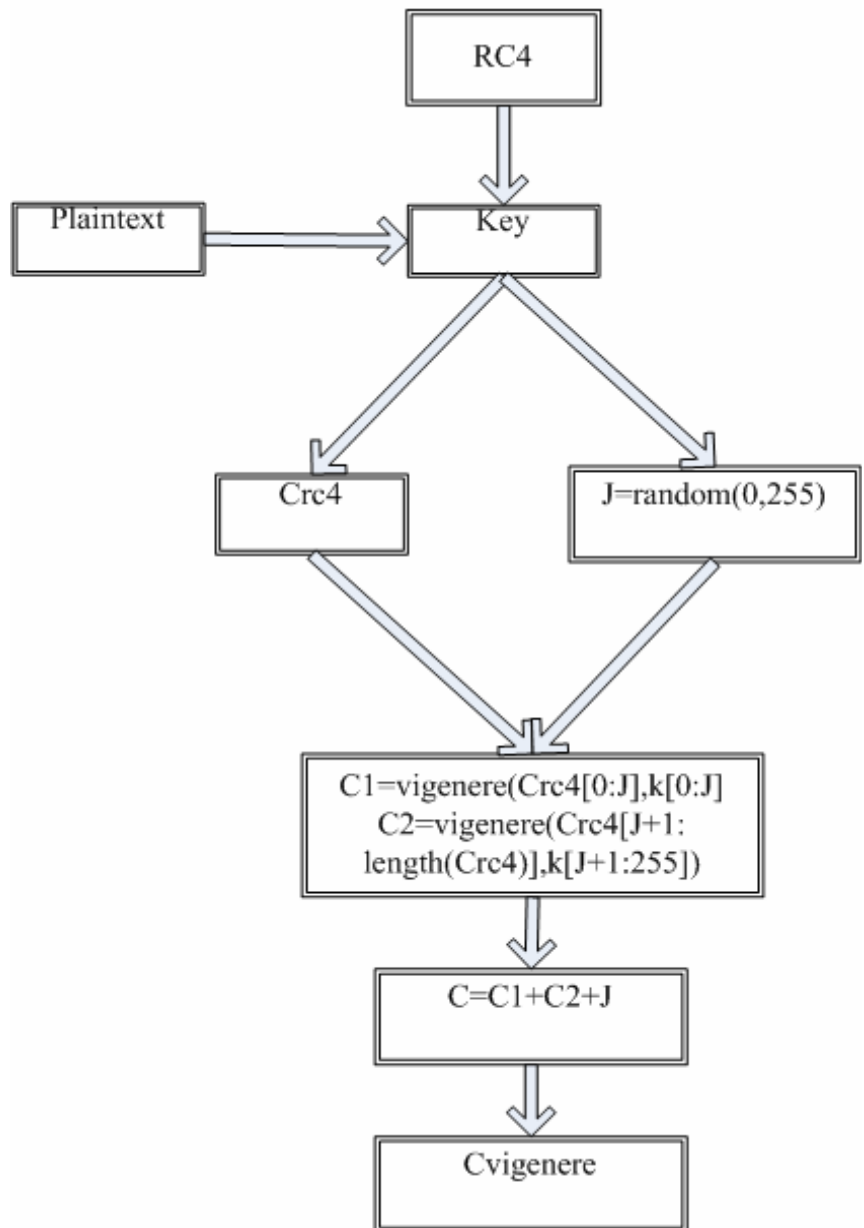

**Figure 8: Encryption process**

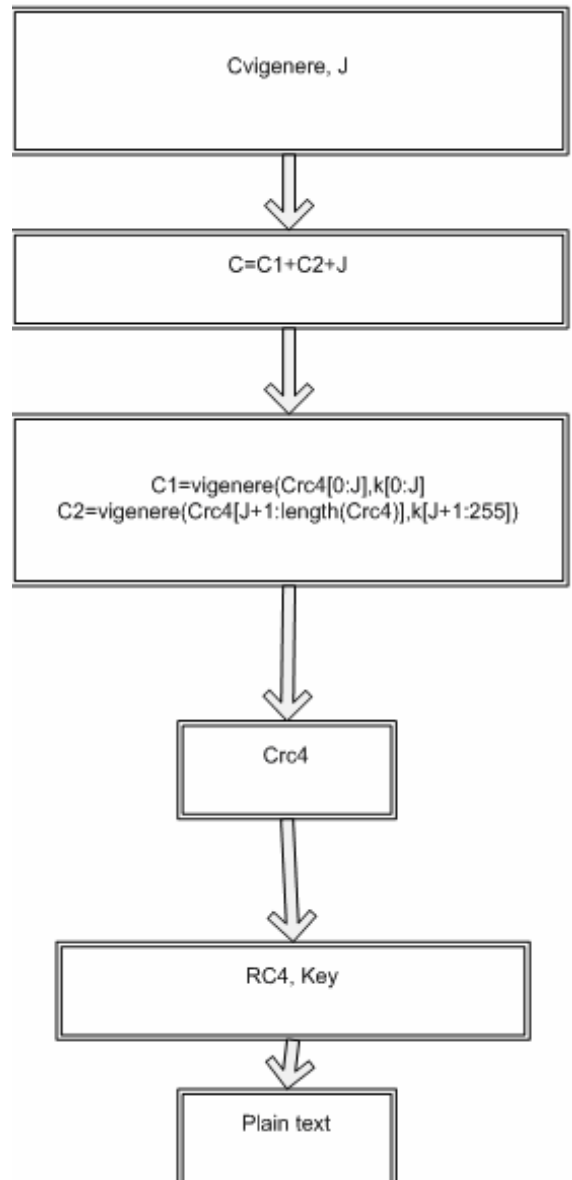

**Figure 9: Decryption process**

**Figure 10: Proposed algorithm VRC4**

The new algorithm is more secure, since, it proofed depending on the theory of the Des that make it more secure by using 3Des or by repeating the same algorithm 3 times, so, we used the same technique by merging RC4 with Vigenère and dividing the key with the cipher to make a new strong algorithm that take long time to be decrypted than the previous one which is RC4.

## 6. Testing and Evaluation

In this section, we show the performance and the security of our algorithm by comparing the Encryption/Decryption time (table 1) of RC4 and VRC4 and the cracking time of the cipher text using the well known open source software for cracking "KisMac" [15, 16] (table 2).

Table 1: Encryption/Decryption time of RC4 versus VRC4

| File size in KB | RC4 Encryption/ decryption time(s) | VRC4 Encryption/ decryption time(s) |
|---|---|---|
| 500 | 42 | 42 |
| 1024 | 154 | 187 |
| 1500 | 240 | 293 |

Table 1 shows the execution time of Encryption/Decryption time of RC4 versus VRC4. The performance of the new algorithm is clear, the time of VRC4 is very close to the time of RC4, i.e., the time factor is well respected.

Table 2: Cracking time of RC4 versus VRC4

| Key composition | RC4 cracked time(s) | VRC4 cracked time(s) |
|---|---|---|
| abcd | 600 | 36000 |
| Qwerty1234 | 1800 | 10000 |
| Qwerty_1234@ | 4500 | Cracking failed |

Table 2 proofs the enhancement on the security of RC4 by using VRC4.

## 7. Conclusion and Future Research

In this paper, we combine two ciphering algorithm RC4 and Vigenère to improve the security of the RC4 algorithm. The simulation shows that the new algorithm VRC4 is more secure than RC4 with same performance. In the future works, we can consider another algorithm to be combined with RC4 or to improve RC4 itself by enhanced the generating key process.

## References
[1] Wasser, S. and Bellere, M., (2001), lecture Notes on cryptography, Cambridge, USA.
[2] Smith, D., (1943), "Substitution Ciphers". Cryptography the Science of Secret Writing: The Science of Secret Writing. Dover Publications. ISBN 048620247X.
[3] David, K. (1999). "On the Origin of a Species". The Codebreakers: The Story of Secret Writing. Simon & Schuster. ISBN 0684831309.
[4] Knudsen, R. (1998). "Block Ciphers, a survey". in Bart Preneel and Vincent Rijmen. State of the Art in Applied Cryptography: Course on Computer Security and Industrial Cryptograph Leuven Belgium, June 1997 Revised Lectures. Berlin ; London: Springer. pp. 29. ISBN 3540654747.
[5] Robshaw, B., (1995), Stream Ciphers Technical Report TR-701, version 2.0, RSA Laboratories.
[6] http://en.wikipedia.org/wiki/RC4_ (cipher)
[7] Stallings, W., (2005) Cryptography and Network Security Principles and Practices, Fourth Edition, New York, USA., 2005.
[8] Souradyuti, P., and Preneel, B., (2004), A New Weakness in the RC4 Keystream Generator and an Approach to Improve the Security of the Cipher. Fast Software Encryption – FSE, pp245 – 259.
[9] Kudsen, L., et al. (1998), "Analysis Method for Alleged RC4." Proceedings, ASIACRYPT '98.
[10] Mister, S., and Tavares, S., (1998) "Cryptanalysis of RC4-Like Ciphers." Proceedings, Workshop in Selected Areas of Cryptography, SAC' 98.
[11] Fluhrer, S., and McGrew, D., (2000) "Statistical Analysis of the Alleged RC4 Key Stream Generator." Proceedings, Fast Software Encryption 2000.

[12] Mantin, I., Shamir, A., (2001) "A Practical Attack on Broadcast RC4." Proceedings, Fast Software Encryption.
[13] Singh, S., (1999). "Chapter 2: Le Chiffre Indéchiffrable". The Code Book. Anchor Book, Random House. pp. 63−78. ISBN 0-385-49532-3.
[14] http://en.wikipedia.org/wiki/Vigen%C3%A8re_cipher
[15] "Under What License is KisMAC Published?". kismac-ng.org. http://kismac-ng.org/wiki/doku.php?id=faq#under_what_license_is_kismac_published. Retrieved 2008-02-22.
[16] http://en.wikipedia.org/wiki/KisMAC